\documentstyle[epsfig,twocolumn,aps,prd]{revtex}
\def\Tr   {{\rm Tr}}
\def\Re   {{\rm Re}}

\def\Ps   {\rlap/P}

\def\Rs   {\rlap/R}
\def\Ls   {\rlap/L}
\def\Los  {\rlap/L_1}
\def\Lts  {\rlap/L_2}

\def\Kps  {\rlap/K_P}
\def\Krs  {\rlap/K_R}

\def\(    {\left(}    \def\)   {\right)}
\def\[    {\left[}    \def\]   {\right]}
\begin{document}
\draft
\title{Infrared and light-cone limit in hot $QED$\thanks{Talk presented 
at the 5th International Workshop on Thermal Field Theories and their 
Applications, Regensburg, Germany, August 1998}}
\author{E. Petitgirard}
\address{Gesellschaft f\"ur Schwerionenforschung mbH, Planckstr. 1,\\
64291 Darmstadt, Germany} 
\maketitle
\begin{abstract}
In hot gauge theories a breakdown of the hard thermal loop expansion
occurs for light-like external momenta or in the infrared region. In
$QED$ where a resummation of ladder diagrams is usually advocated, it is
shown that long range magnetic interations involve a broader set of
graphs. The consequence is a generalized compensation of the
hard modes damping terms at leading order in the infrared limit and near
the light-cone. The relevance of the so-called improved hard
thermal loop resummation scheme is discussed. 
\end{abstract}
\section{Introduction}
\label{sec:intro}

At the scale of the order of the Debye and plasmon masses, gauge
theories at high temperature can no longer be adequately described by
ordinary thermal perturbation theory. A consistent approach requires the
so-called hard thermal loop expansion \cite{r8,klim,wel,brat,wong,tay}. 
Although successful in resolving
many paradoxes, there still remain some fundamental problems in certain
limits. A particular problem is that of the damping rate of a fast
fermion, where self-consistent schemes outside of the hard thermal
loop expansion have been used \cite{smilga,piz}. Another class of 
problems arises
whenever soft external momenta are light-like. This is the case with the
soft photon production rate estimates \cite{baier,brems}. Basically,
like the hard thermal loops themselves, the polarization tensor
involved in the calculation of such processes develops collinear 
singularities. In the same context, a
resummation of asymptotic thermal masses for hard modes, the so-called
improved hard thermal loop scheme, leaves the gauge invariance of the
effective action intact while screening the divergences 
\cite{flech,kraemmer}. 

But in these cases, another candidate for the removal of the collinear
singularities would be the damping term for hard modes. This was
suggested for example in another estimate of the soft photon
production rate \cite{nie}. Indeed, through the interactions with soft modes,
the damping is anomalously large $\gamma\sim g^2T\ln (1/g)$. However if 
$\gamma$ has to be taken into account, modifying only the propagators 
violates the Ward identities. In Abelian theories the latter give
$Q^{\mu}\Pi_{\mu\nu}=0$, but it turns out that
\begin{equation}
Q^{\nu}\Pi_{0\nu}(q_0,0)=\frac{2i\gamma q_0}{q_0+2i\gamma}\frac{e^2T^2}{3}
\end{equation}
Vertex corrections are necessary to restore gauge invariance and leads
to a ladder resummation. The latter arise for instance in the context of
the eikonal expansion of gauge theories \cite{cornwall,hou}. In the 
infrared limit of the
polarization tensor in hot $QED$ \cite{smilga}, using a constant
damping, it was found
that  the damping term gets eliminated from all the components of
$\Pi^{\mu\nu}$. Subsequent works \cite{kraemmer,car} in scalar $QED$ have
reached the same
conclusion for specific limits. Recently a general way to get the
cancellation of the damping term, via simple algebraic compensations,
has been put forward \cite{ckpet,cko}. Besides the non-transversality 
of the polarization
tensor mentioned above, there is another argument concerning gauge
invariance if the case of a non-constant damping is considered. It is
known that when evaluating the damping rate, keeping the external line on
mass-shell with a finite infrared cutoff leads to a gauge independent
contribution \cite{reb}. But with fermion propagators related to
specific internal lines of the
polarization tensor, the integration over the real axis usually leaves
the momentum off mass-shell. For hard fermions there are gauge dependent
pieces off mass-shell besides the factor $e^2T\ln(1/e)$, at least when
going beyond a simple logarithmic approximation. A general compensation
of the damping term seems therefore necessary. 

These considerations do not concern directly the estimates of the soft
photon production rate. The definition of the latter (in covariant
gauges) involves the trace $\Pi^{\mu}_{\mu}$. This term is subleading
compared to the order of the hard thermal loops ($HTL$) and requires more
involved calculations. The approximations justified at the $HTL$ level 
and which are presented in the following are no
longer valid in this case. Nevertheless the soft photon rate
calculations share basic features with the current analysis. The fact
that ladders contribute, both in the infrared and the light-cone limit,
at the same leading order as the $HTL$ is due to a resonance of
denominators \cite{smilga}. The same resonance of denominators occurs 
for the soft
photon rate and leads to a mechanism of enhancement when regularizing 
the collinear divergences with the fermion asymptotic mass
\cite{brems}. The bremsstrahlung processes dominate over previous 
estimates based on the soft fermion loop computations \cite{baier}. 
Now the bremsstrahlung diagrams can be
seen as a first step towards a resummation of higher order graphs. This
is highly expected, from the considerations above concerning
$\Pi^{\mu\nu}$ at leading order on the light-cone, and also from 
semi-classical calculations taken into account the
Landau-Pomeranchuk-Migdal effect \cite{cley}. Finally the following
study about the cancellation of damping terms casts doubt concerning 
attempts to regularize the photon rate in the manner proposed in 
Ref.~\cite{nie}, even if a direct investigation remains to be done. 

Therefore, already at the order of the $HTL$, an important motivation is
to make explicit the equivalence between the infrared and the light-cone
limit. In that respect there are two technical new points compared to
previous studies about ladders. The first one is that Ref.~\cite{ckpet}
left ambiguities with an ill-defined expression for the vertex solution
of the Bethe-Salpeter equation. Such ambiguities prevented a rigorous 
cancellation of damping terms. This problem can be overcome by
reproducing a structure similar to the Landau damping contributions at 
the $HTL$ level, {\it i.e.} a separation between a positive energy and 
a negative energy part. The second point is that diagrammatically, in
order to consider a compensation of a non-constant damping rate, it is 
necessary to go beyond a simple ladder resummation by taking into
account at leading 
order a more general set of graphs. Thus in the infrared region the usual
$HTL$ term will be recovered, as it was already expected from arguments 
using simplified models (constant damping) and using also the kinetic 
approach \cite{jeon,son}. Another motivation is to check the validity of
the so-called improved $HTL$ resummation scheme on the light-cone. 

\section{Leading graphs}
\label{sec:lead}

Several works have been devoted to the cancellation of ladder graphs in
an effective expansion. It is therefore not necessary to enter into
details on this topic. But it is worth recalling, even very briefly, the
power counting argument which leads to a ladder resummation. The
retarded/ advanced formalism \cite{aurenche} is used, since the
structure of Green functions as tree-like diagrams can easily be
seen. However all the calculations could be performed using a different
real time formalism, for example the Schwinger-Keldysh technique 
\cite{keld}. An important point is that leading terms are given by trees
associated with cut internal lines for the photons. For the simple
one-loop $ee\gamma$ vertex where $Q$, $P$ and $R=P+Q$ are respectively
the photon momentum and the incoming and outgoing fermion momenta, and
the particular example of an exchanged transverse photon of momentum
$L$, splitting the product of fermion propagators
$\Delta(P+L)\Delta(R+L)$ along the ways describes in Ref.~\cite{ckpet}
gives  
\begin{eqnarray}
\label{premier}
\tilde{V} & & ^{\mu}_{RAR}(P,Q,-R) = -e^2\int\frac{d^4L}{(2\pi)^4}
 P^t_{\rho\sigma}(L) \gamma^{\rho}(\Rs +\Ls )\gamma^{\mu}\nonumber\\&
&(\Ps +\Ls )\gamma^{\sigma} n(l_0)\rho_T(L)\frac{1}{2Q.(P+L)+Q^2}
\( \Delta_A(R+L)\right.\nonumber\\&
&\left. -\Delta_R(P+L)\) .
\end{eqnarray} 
An estimate of the expression above can be done if the denominator
$2Q.(P+L)+Q^2$ is of the same order as $2P.Q+Q^2$
\begin{eqnarray}
\tilde{V} & & ^{\mu}_{RAR}(P,Q,-R)\sim -e^2\frac{2P^{\mu}+Q^{\mu}}{2Q.P+Q^2}
\int\frac{d^4L}{(2\pi)^4}
 P^t_{\rho\sigma}(L) \nonumber\\&
&\gamma^{\rho}\Ps\gamma^{\sigma} n(l_0)\rho_T(L)
\( \Delta_A(R+L)-\Delta_R(P+L)\) .
\end{eqnarray} 
Under what circumstances this approximation can actually be made and
the denominator extracted from the integral remains to be seen. The
difference of the self-energies written above has the same order as the
damping rate, namely $O(e^2T)$. This is compensated by the factor
$1/e^2$ from the term $1/(2P.Q+Q^2)$. The vertex is therefore of the
same order as its tree-level counterpart and this implies a resummation 
of ladder diagrams 
\begin{figure}[hbt]
\centerline{\epsfig{figure=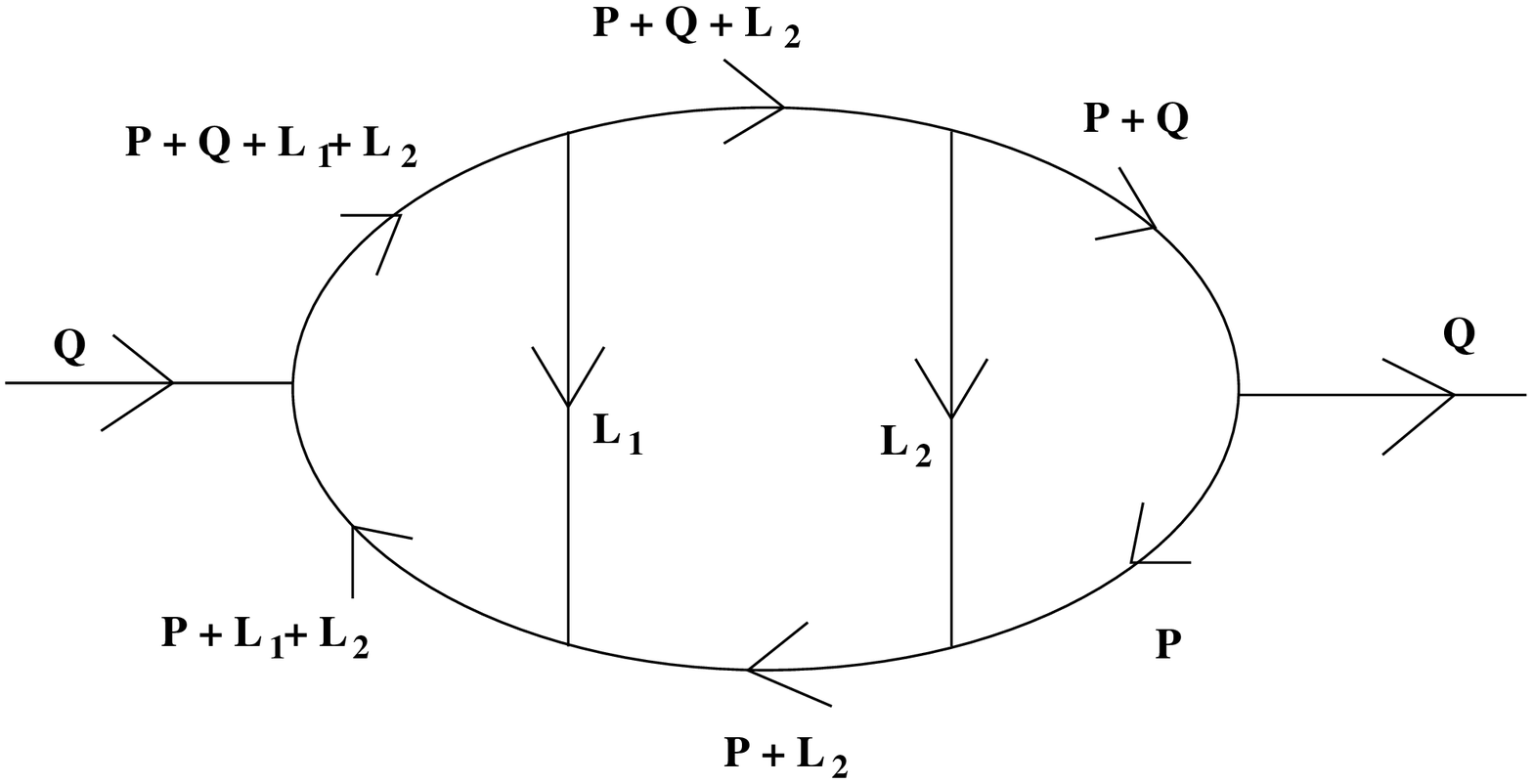,height=30mm}}
\end{figure}
\noindent
with the exchanges of both transverse and longitudinal soft photons. 
The reasoning above was made with bare fermion propagators, but when
a hard electron or positron is close to its mass-shell, a resummation of
its self-energy is necessary. The latter involves one-loop contributions
with a bare fermion propagator for the longitudinal photon case and a
resummed fermion propagator for the transverse photon exchange.

It can be shown that the ladders cancel against the corresponding
self-energies without vertex correction. This is due to a mechanism 
related to the Ward identities. But with the Landau damping part of the
transverse photon spectral density and the absence of Debye screening,
the momenta can reach the very soft scale $O(e^2T)$. This may change
the estimate of multi-loop diagrams. If self-energies with vertex
corrections contribute at leading order as well, as it was pointed out
in Ref.~\cite{blaiz}, then graphs other than ladders might be relevant
for consistency and it is necessary to go beyond what has been done in
previous works \cite{smilga,kraemmer,ckpet,cko}. An estimate of the
multi-loop graphs can be made using the simplified transverse spectral
density introduced in Ref.~\cite{blaiz} 
\begin{equation}
\label{approspectral}
\frac{\rho_T(Q)}{q_0} = \frac{2\pi}{q^2}\delta(q_0) .
\end{equation}
An evaluation of the imaginary part of the two-loop self-energy, with
$P$ on-shell, yields
\begin{equation}
\frac{1}{4p_0}\Tr \( \Ps \Sigma_{RR}(P)\)
= (e^2T)^2\frac{10i}{3(2\pi)^3}\frac{1}{\mu},
\end{equation}
where $\mu$ is a lower cutoff of order $O(e^2T)$.
This result can also be derived within a more complete and rigorous
analysis \cite{pet1} using well known sum rules (see for instance 
Refs.~\cite{braay,piz}). It justifies the fact that, within a pure
logarithmic accuracy, such graphs were discarded \cite{smilga}, but
become necessary when going beyond this approximation in order to 
consider a general (non-constant) damping term. 

For consistency this last remark implies that the vertex for a photon in
the infrared region or near the light-cone must contain at least, not 
only ladders, but ladders with vertex corrections. It turns out that
non-planar diagrams like the crossed graph
\begin{figure}[hbt]
\centerline{\epsfig{figure=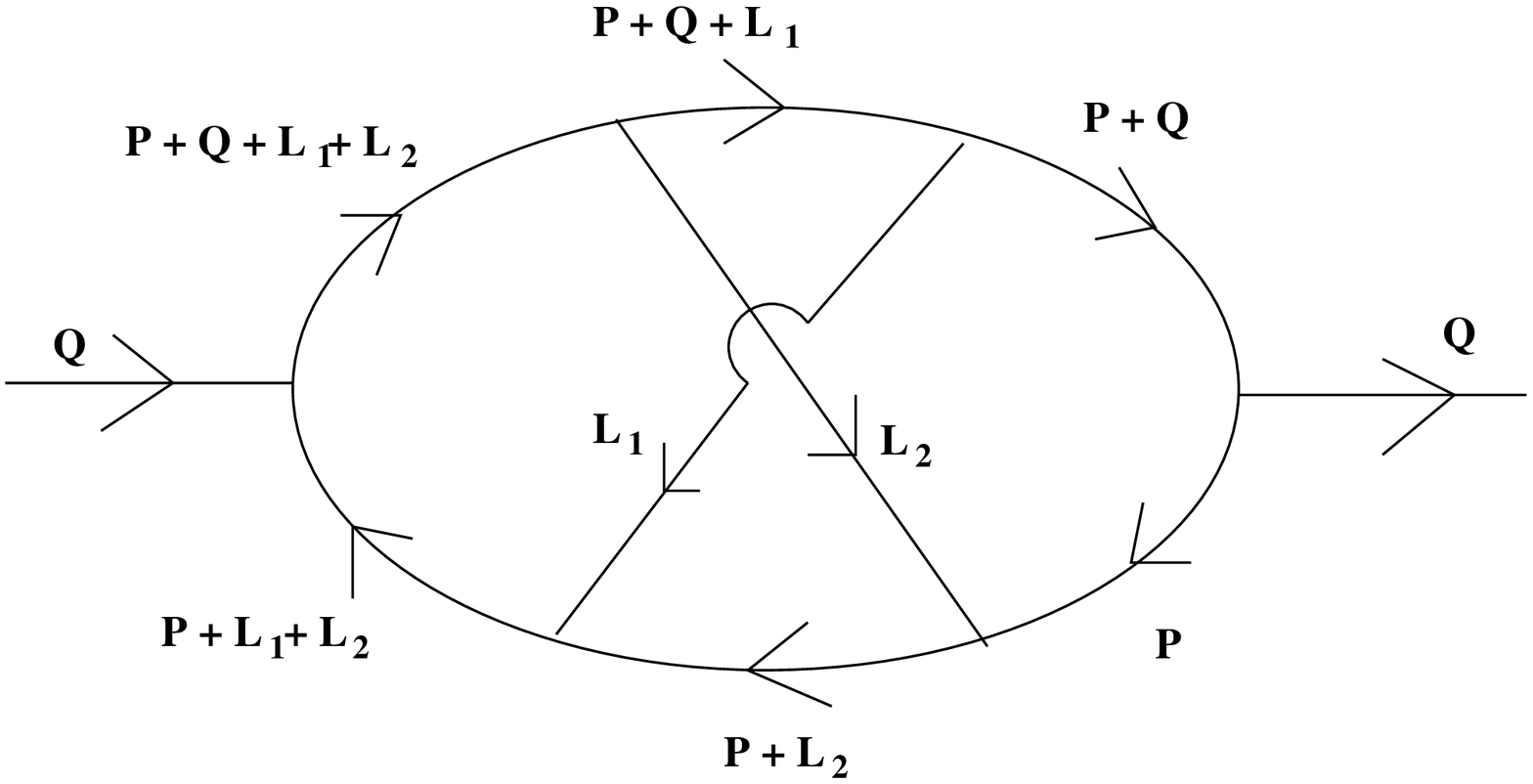,height=30mm}}
\end{figure}
\noindent
contribute also at the same leading order as the tree-level vertex. The 
crossed diagram is of the same order as the quantity
\begin{eqnarray}
\tilde{V}^{+^{\mu}}_{RAR}(P,Q,-R)\sim\frac{\hat{P}^{\mu}}
{q_0+\Omega_P-\Omega_R} & & 
(e^2T)^2\frac{2}{(2\pi)^3}\nonumber\\&
&\times\frac{1}{\mu}\ln\( \frac{\omega_p}{\mu}\) ,
\end{eqnarray}
and is therefore not suppressed by an extra factor $e^2$ contrary to
what has been claimed in Refs.~\cite{ckpet,cko}. In this estimate $\mu$
and the plasma frequency $\omega_p$ are respectively the lower (of 
order $O(e^2T)$) and upper (of order $O(eT)$) cutoffs, $\Omega_P$ and
$\Omega_R$ the moduli of $\vec{p}$ and $\vec{r}$. 

As for graphs with hard photon exchanges non-trivial compensations
between the trace and some denominators kill the mechanism of
enhancement encountered before. Therefore only diagrams with soft
transverse photons have to be added and their resummation along with the
ladder expansion is summarized in the Bethe-Salpeter equation with an 
infinite kernel. The latter is the sum  of one ladder, the subset of 
'two-loop' graphs, 'three-loop' graphs etc... Fermion propagators are 
resummed and, besides
the 'one-loop' diagrams mentioned above, a 'two-loop' vertex correction 
with transverse photons, 'three-loop', etc...must be added. This is
consistent with the mechanism of cancellation of self-energies in the
polarization tensor which now has to be investigated.

\section{Resummed diagrams}
\label{sec:resum}

A method for including the relevant diagrams when going beyond the 
logarithmic approximation must now be considered. This is summarized in
the Bethe-Salpeter equation
\begin{eqnarray}
\label{betsal}
\tilde{V}^{\mu}(P,Q,-R) &=& \gamma^{\mu}+\int\frac{d^4P'}{(2\pi)^4}
 K(P,-P',R',-R)\Krs \Rs'\nonumber\\&
&\Delta(R')\tilde{V}^{\mu}(P',Q,-R')\Ps'\Delta(P')\Kps ,
\end{eqnarray}
where
\begin{eqnarray}
K(P,-P', & & R',-R)\Kps \cdot \Krs = K^1(P,-P',R',-R)\nonumber\\&
&\times\Kps^1 \cdot \Krs^1
+K^2(P,-P',R',-R)\Kps^2 \cdot \Krs^2 +...\nonumber\\&
&R=P+Q;\qquad R'=P'+Q,\nonumber 
\end{eqnarray}
which involves an infinite kernel of four-fermion amplitudes that cannot
be disconnected by cutting two fermion lines (to avoid
overcounting). The first term of this equation represents the simple
ladder resummation with resummed fermion propagators and soft
longitudinal and transverse photons. The second term contains the
summation of the 'two-loop' graphs, namely the crossed diagram and the 
two ladders with one vertex correction. These diagrams involve only
(very) soft transverse photons, as the absence of Debye screening allows
to reach the scale $O(e^2T)$ which compensates the extra factor $e^2$ 
when adding a further loop. The subsequent contributions describe the 
resummation of the "three-loop', 'four-loop' graphs, etc...

Now the resummed fermion propagators involves the self-energy at leading
order when $P^2$ is close to the mass-shell. First this leading order
contribution includes the {\it hard thermal loop} term with the
effective (and constant) asymptotic mass for a hard fermion
\begin{eqnarray}
 & & \Sigma_{HTL}(P)=ap_0\gamma_0+b\vec{p}.\vec{\gamma},\nonumber\\
 & & m_{\infty}^2=2(p_0^2a+p^2b)=\frac{1}{4}e^2T^2.
\end{eqnarray}
Second it must contain also all the exchanges with soft photons. It is
then useful to define the quantity
\begin{equation}
\sigma_R(P)=-\frac{1}{4p_0}\Tr \( \Ps \Sigma_R(P)\) ,
\end{equation}
the imaginary part of which just corresponds to the (general and 
non-constant) damping rate. The propagator reads
\begin{eqnarray}
S_R(P) &=& \frac{i}{\Ps -\Sigma_{HTL}(P)-\Sigma_R(P)}\nonumber\\
 &=& \frac{i\Ps +O(e^2)}{P^2-m_{\infty}^2+2p_0\sigma_R(P)}.
\end{eqnarray}
Discarding the damping term would allow to recover the form advocated in
ref.~\cite{flech}. An important starting point is to decompose the 
resummed propagator into a positive energy part and its negative 
counterpart
\begin{eqnarray}
\Delta_{R/A}(P) &=& \frac{i}{P^2-m_{\infty}^2+2p_0\sigma_{R/A}(P)}
\nonumber\\
&\simeq & \frac{i}{2p}\( \frac{1}{p_0-\Omega_P+\sigma_{R/A}^+(P)}
\right.\nonumber\\&
&\qquad\qquad\left.-\frac{1}{p_0+\Omega_P+\sigma_{R/A}^-(P)}\) \nonumber\\
&=& \Delta^+_{R/A}(P)+\Delta^-_{R/A}(P),
\end{eqnarray}
where $\Omega_P=[\vec{p}^2+m_{\infty}^2]^{1/2}$. Keeping in mind this 
canonical decomposition, an estimate \cite{pet1} enables to guess the 
expression of an {\it improved} vertex   
\begin{eqnarray}
\label{completeform}
\tilde{V} & & _{RAR}^{\mu}(P,Q,-R)=\gamma^{\mu}+
\frac{v^{\mu}}{q_0+\Omega_P-\Omega_R}\( \Sigma^+_R(P)\right.\nonumber\\&
&\left.-\Sigma^+_A(R)\)
+\frac{\bar{v}^{\mu}}{q_0+\Omega_R-\Omega_P}\( 
\Sigma^-_R(P)-\Sigma^-_A(R)\) ,
\end{eqnarray}
where new vectors can be defined
\begin{eqnarray}
v^{\mu} &=& \( 1,\( 1-\frac{m^2_{\infty}}{2p^2}\) \hat{p}^i\) ,\nonumber\\
\bar{v}^{\mu} &=& \( 1,-\( 1-\frac{m^2_{\infty}}{2p^2}\) \hat{p}^i\) , 
\end{eqnarray}
in order for the above vertex to satisfy Ward identities. These vectors
are equal to the unit vectors $(1,\hat{p}^i)$ and $(1,-\hat{p}^i)$ plus
soft contributions unimportant in the determination of the leading order
part of the polarization tensor.

It is now necessary to check that for very soft external photons, this 
vertex is indeed the solution of the Bethe-Salpeter solution, even
though more complicated diagrams than the ladders have been included. It
is also important to see to what extent it is still a solution of 
(\ref{betsal}) in the light-cone limit.

\section{Ladder resummation}
\label{sec:ladder}

The simple ladder resummation has been extensively considered in the
litterature. Here the additional input is the separation of the
expression of the vertex between a positive energy part and its negative
counterpart. This results from the canonical decomposition of the
resummed fermion propagators and due to the basic assumption of an
energetic fermion emitting or absorbing soft photons. Sticking the
expression (\ref{completeform}) into the first term of the Bethe-Salpeter
equation yields
\begin{eqnarray}
\label{oneladd}
\tilde{V} & & ^{1^{\mu}}_{RAR}(P,Q,-R) = \sum_{i=t,l}(-e^2)
\int[dL]^i_{\rho\sigma}\Delta_R(P+L)\nonumber\\&
&\Delta_A(R+L)\gamma^{\rho}(\Rs +\Ls )\tilde{V}^{\mu}_{RAR}(P+L,Q,-R-L) 
\nonumber\\&
&(\Ps +\Ls )\gamma^{\sigma},
\end{eqnarray}
where the notation $[dL]$ for either a transverse or a longitudinal 
spectral density stands for
\begin{equation}
[dL]^i_{\rho\sigma}=\frac{d^4L}{(2\pi)^4}n(l_0)\rho_i(L)
P^i_{\rho\sigma}(L).
\end{equation}
Contracting the spinors gives
\begin{eqnarray}
(\Rs +\Ls )\gamma^{\mu}(\Ps +\Ls ) &\simeq & 2P^{\mu}\Ps ,\nonumber\\
(\Rs +\Ls )v^{\mu}\Sigma^+(P+L)(\Ps +\Ls ) 
&\simeq & -2P^{\mu}\sigma^+(P)\Ps .
\end{eqnarray}
Soft terms may be discarded and the basic picture of an energetic
electron (or positron) undergoing interactions with soft photons (momenta
of order $O(eT)$ at most) implies that the former is always close to its
mass-shell ($P^2\sim O(eT^2)$ at most). This allows to make several
approximations in the equations above. The expression (\ref{oneladd})
now reads
\begin{eqnarray}
\label{intermvert}
\tilde{V} & & ^{1^{\mu}}_{RAR}(P,Q,-R)=\sum_{i=t,l}(-e^2)
\int[dL]^i_{\rho\sigma}
\Delta_R(P+L)\nonumber\\&
&\Delta_A(R+L)2P^{\mu}\gamma^{\rho}\Ps \gamma^{\sigma}
\[ 1-\frac{\sigma_R^+(P+L)-\sigma_A^+(R+L)}{q_0+
\Omega_{P+L}-\Omega_{R+L}}\right.\nonumber\\&
&\left.-\frac{\sigma_R^-(P+L)-\sigma_A^-(R+L)}{q_0+\Omega_{R+L}-
\Omega_{P+L}}\] .
\end{eqnarray}
This enables also to split the product of propagators
$\Delta(P)\Delta(R)$ according to their canonical decompositions (mixed
positive/ negative energy terms are shown to be subleading)
\begin{eqnarray}
\Delta_R(P) & & \Delta_A(R)\simeq \frac{i}{2p}\( \frac{\Delta^+_R(P)
-\Delta^+_A(R)}{q_0+\Omega_P-\Omega_R-\sigma_R^+(P)+\sigma_A^+(R)} 
 \right.\nonumber\\&
&\left.- \frac{\Delta^-_R(P)-\Delta^-_A(R)}{q_0+\Omega_R-\Omega_P
+\sigma_A^-(R)-\sigma_R^-(P)} \) , 
\end{eqnarray}
and to relate the positive or negative energy parts of the propagators
to the corresponding contributions of the vertex 
\begin{eqnarray}
& & \[ \Delta^+_R(P)- \Delta^+_A(R)\] \[ 1-\frac{\sigma_R^+(P)-
\sigma_A^+(R)}{q_0+\Omega_P-\Omega_R}\right.\nonumber\\&
&\left.\quad-\frac{\sigma_R^-(P)-\sigma_A^-(R)}{q_0+\Omega_R-\Omega_P}\]
\nonumber\\&
&\quad\simeq\[ \Delta^+_R(P)-\Delta^+_A(R)\] \[ 1-\frac{\sigma_R^+(P)-
\sigma_A^+(R)}{q_0+\Omega_P-\Omega_R}\] .
\end{eqnarray}
It can be easily seen that the algebraic cancellation occurs quite
naturally using this decomposition. The first term of the Bethe-Salpeter
equation equation may now be written
\begin{eqnarray}
\tilde{V}^{1^{\mu}}_{RAR} & & (P,Q,-R)=\sum_{i=t,l}(-ie^2)
\int[dL]^i_{\rho\sigma}
\gamma^{\rho}\Ps \gamma^{\sigma}\nonumber\\&
&\( v^{\mu}\frac{\Delta^+_R(P+L)-\Delta^+_A(R+L)}{q_0+\Omega_{P+L}
-\Omega_{R+L}} \right.\nonumber\\&
&\left.\qquad+\bar{v}^{\mu}\frac{\Delta^-_R(P+L)-\Delta^-_A(R+L)}
{q_0+\Omega_{R+L}-\Omega_{P+L}}\) . 
\end{eqnarray}
It is now crucial to see to what extent the denominators without damping
can be extracted from the integral. Expanding their expressions with the
soft terms from $L$ and $Q$ gives for example
\begin{eqnarray}
\label{expansion}
q_0+\Omega_{P+L}-\Omega_{R+L} &=& q_0+\Omega_P-\Omega_R \nonumber\\&
&-\frac{\vec{q}}{p}.\[ \vec{l}-\hat{p}(\hat{p}.\vec{l})\] +O(e^3T). 
\end{eqnarray}  
In the infrared limit ($q_0,q\sim O(e^2T)$) and on the light-cone for a
not too collinear configuration ($\hat{p}.\hat{q}\sim\pm1+O(e)$) the
terms following $q_0+\Omega_P-\Omega_R$ may always be neglected. This
is obvious in the infrared region. For the light-cone limit it turns
out that $\hat{q}\sim\hat{p}$  and consequently 
$(\vec{q}/p).( \vec{l}-\hat{p}(\hat{p}.\vec{l}))\sim O(e^3T)$. However
one of the main points is to notice that in a more collinear
configuration these subsequent contributions are no longer
neglegible. Indeed in this last case $q_0+\Omega_P-\Omega_R$ is of the
same order $O(e^3T)$. This is the situation where the effect of the
asymptotic mass is important. Therefore, outside this limit, {\it i.e.}
in the infrared and in a 'weak' light-cone limit, the relevant
denominators may be extracted and the expression of the vertex 
finally gives
\begin{eqnarray}
\tilde{V} & & _{RAR}^{1^{\mu}}(P,Q,-R)=
\frac{v^{\mu}}{q_0+\Omega_P-\Omega_R}\(
\Sigma^{1^+}_R(P)-\Sigma^{1^+}_A(R)\) \nonumber\\&
&+\frac{\bar{v}^{\mu}}{q_0+\Omega_R-\Omega_P}\( 
\Sigma^{1^-}_R(P)-\Sigma^{1^-}_A(R)\) .
\end{eqnarray}
But of course this part does not correspond to the the full contribution
of the vertex (\ref{completeform}) since it is necessary to go beyond a
simple ladder resummation. 

\section{N-loop diagrams resummation}
\label{sec:nloop}

So far only a simple ladder expansion has been considered. Since also
'higher-order' graphs have been shown to contribute at leading order it
is therefore necessary to take into account their resummation. This
corresponds to an infinite kernel for the Bethe-Salpeter equation. 

As a first step beyond the ladders, the algebraic cancellation of
self-energies for the 'two-loop' graphs must now be considered. Inserting
the vertex (\ref{completeform}) gives for the 'two-loop' crossed diagram
\begin{eqnarray}
\tilde{V} & & ^{2^{\mu}}_{Cr_{RAR}}(P,Q,-R) = e^4\int[dL_1]^t_{\alpha\beta}
[dL_2]^t_{\sigma\rho}\Delta_A(R+L_1)\nonumber\\&
&\Delta_A(R+L_1+L_2)\Delta_R(P+L_1+L_2)\Delta_R(P+L_2)Sp(R)\nonumber\\&
&\tilde{V}^{\mu}_{RAR}(P+L_1+L_2,Q,-R-L_1-L_2) Sp(P),
\end{eqnarray} 
where the $\Delta$'s are the resummed fermion propagators and 
\begin{eqnarray}
Sp(R) &=&\gamma^{\alpha}(\Rs +\Los )
\gamma^{\sigma}(\Rs +\Los +\Lts ),
\nonumber\\
Sp(P) &=& (\Ps +\Los +\Lts )\gamma^{\beta}(\Ps +\Lts )\gamma^{\rho}.
\end{eqnarray}
The product of propagators $\Delta(P+L_1+L_2)\Delta(R+L_1+L_2)$ attached
to the internal vertex has to be split in the same way as for the
ladders. The decomposition of fermion propagators into positive and
negative energy parts allows to separate a positive energy contribution
from its negative counterpart in the vertex. The algebraic cancellation
still occurs and is expressed through the replacements of 
$(q_0\pm\Omega_{P+L_1+L_2}\mp\Omega_{R+L_1+L_2}\pm\sigma_{R_{P+L_1+L_2}}
\mp\sigma_{A_{R+L_1+L_2}})$ by the denominators
$(q_0\pm\Omega_{P+L_1+L_2}\mp\Omega_{R+L_1+L_2})$. Finally in the
infrared and in a weak light-cone limit the $L_i$-dependent terms can
always be neglected and $(q_0+\Omega_P-\Omega_R)$ or 
$(q_0+\Omega_R-\Omega_P)$ turns out to be a good aproximation for 
$(q_0+\Omega_{P+L_1+L_2}-\Omega_{R+L_1+L_2})$ or 
$(q_0+\Omega_{R+L_1+L_2}-\Omega_{P+L_1+L_2})$. Extracting the first
denominators from the integral yields 
\begin{eqnarray}
\label{crossed}
\tilde{V}& &^{2^{\mu}}_{Cr_{RAR}}(P,Q,-R) = 
\frac{v^{\mu}}{q_0+\Omega_P-\Omega_R}ie^4\int[dL_1]^t_{\alpha\beta}
[dL_2]^t_{\sigma\rho}\nonumber\\&
&Sp(P,R)\Delta^+_A(R+L_1)\Delta^+_R(P+L_2)
\( \Delta^+_R(P+L_1+L_2)\right.\nonumber\\&
&\left.-\Delta^+_A(R+L_1+L_2)\) +n.e. 
\end{eqnarray} 
with
\begin{equation}
Sp(P,R)=\gamma^{\alpha}(\Rs +\Los )\gamma^{\sigma}
(\Ps +\Los +\Lts )\gamma^{\beta}(\Ps +\Lts )\gamma^{\rho},
\end{equation}
(the terms $L_i$ in the numerator are subleading but have been kept to
stress the similarity with Ward identities). If now the other 'two-loop'
diagrams are considered, namely the vertex correction attached to the
soft photon line $L_1$ and its symmetric counterpart, the same procedure
as before can be applied when inserting the full vertex. In particular
the resulting denominators $(q_0\pm\Omega_{P+L_i}\mp\Omega_{R+L_i})$ may
be approximated by $(q_0\pm\Omega_P\mp\Omega_R)$ in the aforementioned
limits. Adding the positive energy and negative energy contributions of 
these diagrams with the crossed graph gives 
\begin{eqnarray}
\tilde{V} & & _{RAR}^{2^{\mu}}(P,Q,-R) = 
\frac{v^{\mu}}{q_0+\Omega_P-\Omega_R}\(
\Sigma^{2^+}_R(P)-\Sigma^{2^+}_A(R)\) \nonumber\\&
& +\frac{\bar{v}^{\mu}}{q_0+\Omega_R-\Omega_P}\( 
\Sigma^{2^-}_R(P)-\Sigma^{2^-}_A(R)\) .
\end{eqnarray} 

The reasoning above can be generalized to the 'N-loop' vertices in the
Bethe-Salpeter equation. These vertices involve resummed fermion
propagators and the exchange of soft transverse photons. They induce
four-fermion amplitudes which cannot be disconnected by cutting two
fermion lines. Each of these graphs have internal legs $P+L_i+...+L_j$
and $R+L_i+...+L_j$ attached to the external photon line via the
complete vertex. Splitting the product of propagators 
$\Delta(P+L_i+...+L_j)\Delta(R+L_i+...+L_j)$ as before leads to the
algebraic cancellation of the self-energies $\sigma$, {\it i.e.} to the
replacement of the common denominator $(q_0+\Omega_{P+L_i+...}-\Omega
_{R+L_i...}+\sigma_{R_{P+L_i+...}}-\sigma_{A_{R+L_i+...}})$ by $(q_0+\Omega_
{P+L_i+...}-\Omega_{R+L_i...})$. Again in the infrared region and near
the light-cone without a too strong collinearity between the photon and
the electron the latter denominator can be approximated by
$q_0+\Omega_P-\Omega_R$ ($q_0+\Omega_R-\Omega_P$ for the negative energy
part) and extracted from the integral. What is left is completely
equivalent to a contraction of each graph with $Q^{\mu}$, times the
vector $v^{\mu}/(q_0+\Omega_P-\Omega_R)$ (and its negative energy 
counterpart $\bar{v}^{\mu}/(q_0+\Omega_R-\Omega_P)$). It is therefore
natural for a demonstration at any order to rely on Ward identities
between vertices with resummed fermions and cut soft transverse
photons and the corresponding self-energies. The other additional input
being the separation between a positive energy contribution and the
negative one. This is due to the basic assumption of an energetic (hard)
fermion interacting with soft photons. Each diagram is related to
specific terms which would be obtained with a simple contraction with
$Q^{\mu}$
\begin{eqnarray}
\tilde{V} & & ^{N^{\mu}}_{M\, K_{RAR}}(P,Q,-R) = \frac{v^{\mu}}
{q_0+\Omega_P-\Omega_R}
\( \Sigma^{N^+}_{M\, K_R}(P)\right.\nonumber\\&
&\left.-\Sigma^{N^+}_{M\, K_A}(R)\) 
+\frac{\bar{v}^{\mu}}{q_0+\Omega_R-\Omega_P}
\( \Sigma^{N^-}_{M\, K_R}(P)\right.\nonumber\\&
&\left.-\Sigma^{N^-}_{M\, K_A}(R)\) ,
\end{eqnarray} 
the sum of which corresponds to a self-energy with a particular vertex
correction
\begin{equation}
\Sigma^{N^{\pm}}_{M_R}(P)=\sum_K\Sigma^{\pm}_{M\, K_R}(P).
\end{equation}
Adding these self-energies yields the 'N-loop' order $\Sigma$
\begin{equation}
\Sigma^{N^{\pm}}_R(P)=\sum_M\Sigma^{\pm}_{M_R}(P),
\end{equation}
and finally the contribution of all 'N-loop' self-energies gives the
complete leading order expression
\begin{equation}
\Sigma^{\pm}_R(P)=\sum_M\Sigma^{N^{\pm}}_R(P),
\end{equation}
which allows to recover the {\it improved} vertex
(\ref{completeform}). The latter is therefore the solution of the
Bethe-Salpeter equation for the particular cases considered so far. 

\section{Cancellation of damping terms}
\label{sec:damp}

In the infrared limit the improved vertex
\begin{eqnarray}
\tilde{V} & & _{RAR}^{\mu}(P,Q,-R)=\gamma^{\mu}+
\frac{v^{\mu}}{q_0+\Omega_P-\Omega_R}\( \Sigma^+_R(P)\right.\nonumber\\&
&\left.-\Sigma^+_A(R)\)
+\frac{\bar{v}^{\mu}}{q_0+\Omega_R-\Omega_P}\( 
\Sigma^-_R(P)-\Sigma^-_A(R)\) ,
\end{eqnarray}
is therefore shown to be the solution of the Bethe-Salpeter equation. In
the light-cone limit where the effects of the asymptotic mass are
relevant, this is not true any more, since approximations valid in
the former case are no longer legitimate. Nevertheless it is always
interesting to look at the expression of $\Pi^{\mu\nu}(Q)$ provided by
this vertex. To deal with a general (non-constant) damping term, the
trick consists in coming back to the imaginary time formalism,
comparisons between quantities like $i\omega_n$, $\Omega_P$ and 
$\sigma(i\omega_n,p)$ being easier. In the $R/A$ formalism, the retarded
part of the polarization tensor can be written as
\begin{eqnarray}
i\Pi & & ^{\mu\nu}_{RR}(Q)= -e^2\Tr \int\frac{d^4P}{(2\pi)^4}\( \Ps 
\gamma^{\mu}\Rs \) 
\Bigg\{ \(
\frac{1}{2}-n_F(p_0)\) \nonumber\\&
&\Delta_R(R)\[ \tilde{V}^{\nu}_{RRA}
 \Delta_R(P)-\tilde{V}^{\nu}_{ARA}
\Delta_A(P)\] 
+\( \frac{1}{2}\right.\nonumber\\&
&-n_F(r_0)\Big) 
\Delta_A(P)\[  \tilde{V}^{\nu}_{ARA} 
\Delta_R(R)-\tilde{V}^{\nu}_{ARR} \Delta_A(R)\] \Bigg\} ,
\nonumber\\
\tilde{V} & & ^{\nu}_{\alpha\beta\delta}
=\gamma^{\nu}+\tilde{V}^{1^{\nu}}_{\alpha\beta\delta}(P,Q,-R).
\end{eqnarray}
Contracting the spinors gives a part sensitive to the infrared or
light-cone limit plus possible tadpole terms. The latter do not involve
vertex and damping corrections, or more precisely it can be shown that
such corrections are subleading. Splitting the products of propagators
yields denominators containing the retarded and advanced damping
terms. These denominators cancel against the numerators of the form 
$1+(\sigma_P-\sigma_R)/(q_0+\Omega_P-\Omega_R)$ derived from the
equation (\ref{completeform}). Taking $\Pi^{00}$ as a particular
example, a simpler expression is thus obtained
\begin{eqnarray}
i\Pi & &^{00}_{RR}(Q)= -4ie^2\int\frac{d^4P}{(2\pi)^4}p\left\{ \(
\frac{1}{2}-n_F(p_0)\) \times\right.\nonumber\\&
&\[ \frac{\delta^+(P)}{q_0+\Omega_P-\Omega_R} 
-\frac{\delta^-(P)}{q_0+\Omega_R-\Omega_P}\] 
-\( \frac{1}{2}-n_F(r_0)\) \nonumber\\&
&\left.\[ \frac{\delta^+(R)}{q_0+\Omega_P
-\Omega_R}
 -\frac{\delta^-(R)}{q_0+\Omega_R-\Omega_P}
\] \right\} ,\nonumber\\
\delta & & ^{\pm}(P)=\Delta^{\pm}_R(P)-\Delta^{\pm}_A(P).         
\end{eqnarray}
The integration over $p_0$ can be converted into a contour integral in
the complex half-plane where there are no discontinuities from the
propagators but only the poles given by the Matsubara frequencies. The
contour can be composed of the real axis, two quarter-circles expanding
at infinity, the contribution of which tends to zero, and a part circling
the upper or lower poles of the Fermi-Dirac factor. The contribution is
finally reduced to the sum of the residues corresponding to the
Matsubara frequencies 
\begin{eqnarray}
\label{matsuform}
i & & \Pi^{00}_{RR}(Q)=
4i\pi T e^2\int\frac{d^3p}{(2\pi)^3}\sum_n
  \[ \frac{1}{q_0+\Omega_P-\Omega_R}\times\right.\nonumber\\& 
&\left.\( \frac{1}{\omega_n-\Omega_P+\sigma^+(\omega_n,p)}-
\frac{1}{\omega_n-\Omega_R+\sigma^+(\omega_n,r)}\)
\right.\nonumber\\&
&\left.-\frac{1}{q_0+\Omega_R-\Omega_P}
\(
\frac{1}{\omega_n+\Omega_P+\sigma^-(\omega_n,p)}\right.\right.\nonumber\\&
&\left.\left.-\frac{1}{\omega_n+\Omega_R+\sigma^-(\omega_n,r)} \) \] .    
\end{eqnarray}
It is mainly the difference $\Omega_P-\Omega_R\sim \hat{p}.\vec{q}$
which matters. Its order of magnitude is $O(eT)$ in the light-cone
limit, and $O(e^2T)$ or below in the infrared region. It is much larger
than $\sigma(i\omega_n,p)-\sigma(i\omega_n,r)$ in each case. The
contribution due to the damping terms is therefore negligible. The
expression given by simple poles in real time is re-established and the
improved $HTL$
\begin{eqnarray}
i\Pi & &^{00}_{RR}(Q)= -4i\pi e^2\int\frac{d^3P}{(2\pi)^4}
\( n_F(\Omega_R)-n_F(\Omega_P)\) \nonumber\\&
&\[ \frac{1}{q_0+\Omega_P-\Omega_R}
+\frac{1}{q_0+\Omega_R-\Omega_P}\] ,
\end{eqnarray}
is recovered. The complex energy $q_0=\Re q_0+i\epsilon$ allows to get
the usual Landau damping part. 

In the infrared limit and on the light-cone when
$\hat{p}.\hat{q}\sim\pm1+O(e)$ the resummation of all the
aforementioned vertex diagrams cancels against the self-energy
insertions. The result is found to be the {\it improved} $HTL$, which
just corresponds to the usual $HTL$ in these cases. 

However in a strong light-cone limit where the effects of the asymptotic
mass are important, the {\it improved} vertex (\ref{completeform}) is no
longer solution of the Bethe-Salpeter equation. The expansion
(\ref{expansion}) is not reduced any more to the first term as it is in 
the first two cases. The resummation scheme advocated in
Ref.~\cite{flech} is therefore not correct, or at least not complete. 
However the cancellation of (\ref{completeform}) with self-energies 
whatever the gauge is, provides a further and deeper justification of 
the gauge invariance of the {\it improved} $HTL$ of Ref.~\cite{flech}. 
But to propose a more complete resummation scheme on the light-cone, 
other ways of dealing with the Bethe-Salpeter equation are necessary 
\cite{pet}.

%


\begin{references}
%
\bibitem{r8} 
R.~D.~Pisarski, Physica {\bf A158}, 146, 246 (1989);
Phys. Rev. Lett. {\bf 63}, 1129 (1989).
\bibitem{klim} V.~V.~Klimov, Sov. Phys. JETP {\bf 55}, 199 (1982).
\bibitem{wel} H.~A.~Weldon, Phys. Rev. {\bf D26}, 1384, 2789 (1982).
\bibitem{brat} E.~Braaten and R.~D.~Pisarski,
Nucl. Phys. {\bf B337}, 569 (1990); {\bf B339}, 310 (1990).
\bibitem{wong} J.~C.~Taylor and S.~M.~H.~Wong,
Nucl. Phys. {\bf B346}, 115 (1990).
\bibitem{tay} J.~Frenkel and J.~C.~Taylor,
 Nucl. Phys. {\bf B334}, 199 (1990); Z. Phys. {\bf C49}, 515 (1991).
\bibitem{smilga} V.~V.~Lebedev and A.~V.~Smilga,
Ann. Phys. (N.Y.) {\bf 202}, 229 (1990); Physica {\bf A181}, 187 (1992).
\bibitem{piz} R.~D.~Pisarski, Phys. Rev. {\bf D47}, 5589 (1993).
\bibitem{baier} R.~Baier, S.~Peign\'e and D.~Schiff, Z. Phys. {\bf C62},
337 (1994); P.~Aurenche, T.~Becherrawy and E.~Petitgirard, {\it preprint
ENSLAPP-A-452-93} (hep-ph/9403320).
\bibitem{brems} P.~Aurenche, F.~Gelis, R.~Kobes and E.~Petitgirard, 
Phys. Rev. {\bf D54}, 5274 (1996); Z. Phys. {\bf C75}, 315 (1997).
\bibitem{flech} F. Flechsig and A. Rebhan, Nucl. Phys. {\bf B464}, 279
(1996).
\bibitem{kraemmer} U.~Kraemmer, A.~K.~Rebhan, and H.~Schulz,
Ann. Phys. (N.Y.) {\bf 238}, 286 (1995).
\bibitem{nie} A.~Ni\'egawa, Phys. Rev. {\bf D56}, 1073 (1997);
Phys. Rev. {\bf D55}, 4997 (1997).
\bibitem{cornwall} J.~M.~Cornwall and G.~Tiktopoulos,
 Phys. Rev. {\bf D15}, 2937 (1977).
\bibitem{hou} J.~M.~Cornwall and W.-S.~Hou, Phys. Lett. {\bf B153}, 173
(1985).
\bibitem{car} M.~Carrington, Phys. Rev. {\bf D48}, 3836 (1993).
\bibitem{ckpet} M.~Carrington, R.~Kobes and E.~Petitgirard,
 Phys. Rev. {\bf D57}, 2631 (1998).
\bibitem{cko} M.~Carrington and R.~Kobes,
 Phys. Rev. {\bf D57}, 6372 (1998).
\bibitem{reb} A.~K.~Rebhan, Phys. Rev. {\bf D46}, 4779 (1992).
\bibitem{cley} J.~Cleymans, V.~V.~Goloviznin and K.~Redlich, Phys. Rev. 
{\bf D47}, 989 (1993); Z. Phys. {\bf C59}, 495 (1993).
\bibitem{jeon} S.~Jeon, Phys. Rev. {\bf D52}, 3591 (1995); S.~Jeon and
 L.~Yaffe, Phys. Rev. {\bf D53}, 5799 (1996).
\bibitem{son} P.~Huet and D.~T.~Son, Phys. Lett. {\bf B393}, 94 (1997).
\bibitem{aurenche} P.~Aurenche and T.~Becherrawy,
Nucl. Phys. {\bf B379}, 259 (1992); M~.A.~van Eijck and C.~G.~van Weert, 
Phys. Lett. {\bf B278}, 305 (1992).
\bibitem{keld} L.~V.~Keldysh, Sov. Phys. JETP {\bf 20}, 1018 (1964); 
See also E.~M.~Lifshitz and L.~P.~Pitaevsky, {\it Physical Kinetics}
(Pergamon Press, Oxford, 1981).
\bibitem{blaiz} J.~P.~Blaizot and E.~Iancu,
Phys. Rev. {\bf D55}, 973 (1997); Phys. Rev. {\bf D56}, 7877 (1997).
\bibitem{pet1} E.~Petitgirard, {\it preprint GSI-98-45}
(hep-ph/9808344), to appear in Phys. Rev. {\bf D}. 
\bibitem{braay} E.~Braaten and T.~C.~Yuan,
Phys. Rev. Lett. {\bf 66}, 2183 (1991).
\bibitem{pet} E.~Petitgirard, in preparation.
%
\end{references}
\end{document}